# A Method for Detecting Life-Threatening Signals in Serum Potassium Level after Myocardial Infarction

Michal Shauly-Aharonov[1,*], Moshe Pollak[1], Ygal Plakht[2]

[1]The Hebrew University of Jerusalem, Mount Scopus, Jerusalem 91905, Israel

[2]Soroka University Medical Center, Beer-Sheva, Southern 84101, Israel

[*]email: mshauly@gmail.com

## SUMMARY

Clinical guidelines recommend maintaining serum potassium levels between 4.0 and 5.0 mEq/L in patients with acute myocardial infarction (MI). These guidelines are based on recent studies that found significant associations between crossing of absolute potassium limits (by in-hospital mean or by min/max values) and mortality. This paper investigates a different approach: we hypothesized that a change in the potassium level may be a harbinger of short survivability, rather than crossing of absolute boundaries. Our objectives were: (1) to examine if a "change in mean" indicator has the ability to distinguish between survivors and non-survivors of MI hospitalization, and if so, (2) to formulate a framework for detecting life-threatening changes in potassium level of patients hospitalized with MI. The study included 195 patients who were hospitalized for MI from 2002 to 2014, with at least 40 potassium measurements (i.e., severely ill). In a retrospective analysis we found evidence that the "change in mean" criterion significantly discriminated between survivors and non-survivors. A threshold for raising an alarm was specified by plotting an ROC curve and choosing the value that yields the best combination of sensitivity and specificity. In this case, the method detected ~80% of the patients that eventually died, while wrongly alerting for only ~40% of the survivors. The proposed approach is not intended for replacing the absolute-level protocols but to add valuable knowledge to cardiologists.

# 1 Introduction

Cardiovascular diseases (CVD) and cardiac death rates have declined significantly between 1990 and 2010 in most high-income countries (Moran et al., 2014), yet CVD continue to claim many lives annually. In 2011, CVD still accounted for one of every three deaths in the USA (Mozaffarian et al., 2015). Extreme values of potassium ($K^+$) in the blood are dangerous to the heart since they can impair the activity of nerves and muscles, and induce cardiac arrhythmia up to heart failure and death, unless treated promptly. However, hyperkalemia (elevated concentration of potassium in the blood) and hypokalemia are usually difficult to diagnose, since the symptoms can be fairly mild and non-specific and may be due to many different health problems. The most important treatment in dyskalemia is diagnosing it as early as possible and addressing the cause.

The risk of arrhythmia and death is much greater after a myocardial infarction (MI; heart attack), where injured heart tissue does not conduct electrical impulses as fast as normal heart tissue. It is therefore important to identify and treat patients with MI who are at high risk for arrhythmia, and to enhance methodologies for prediction of cardiac death (Fishman et al., 2010). Numerous studies have been recently published on the relationship between potassium and mortality in patients with MI; Goyal et al. (2012) investigated this relationship in patients hospitalized with acute MI and found that the lowest in-hospital mortality was in those with mean potassium level between 3.5 and 4.5 mEq/L; Hessels et al. (2015) evaluated the relationship between potassium levels during ICU-stay and in-hospital mortality, and found that hypokalemia (minimal in-hospital value < 3.5 mEq/L), hyperkalemia (maximal value > 5.0 mEq/L) and potassium variability were independently associated with increased mortality; Shiyovich, Gilutz and Plakht (2015) explored the relationship between in-hospital potassium levels and post-discharge long-term mortality among patients with MI, and demonstrated a significant U-shaped relationship with lowest mortality among patients with potassium level between 4.0 and 4.5 mEq/L. However, most prior literature aims to explain mortality by crossing of absolute potassium limits (may it be by the in-hospital mean or by min/max values).

In patients after MI, the American heart association guidelines recommend maintaining potassium levels between 4.0 and 5.0 mEq/L (Goyal et al., 2012). Recent studies argue that in practice dyskalemia is a continuum, thus these guidelines are ineffective (Goyal et al. 2012; Fishman et al., 2010). However, standard protocols in hospitals nowadays monitor only for a crossing of absolute potassium boundaries. A more personalized approach (e.g., monitoring for a change in the individual's mean and/or variability of potassium level) has not been examined yet.

In this paper, we aim to test the hypothesis that a real change in the patient's mean level of potassium is associated with in-hospital mortality after MI. If so, a secondary objective would be to formulate a framework for detecting life-threatening changes in potassium level of patients hospitalized with MI. The "change in mean" approach has not been previously employed in the area of potassium control; it is innovative in two major aspects: first, it detects changes in the individual's pattern of variation over time, rather than crossing an absolute limit, and thus is personalized to each patient; second, it accumulates data by using a risk statistic that considers the whole in-hospital history of the patient, unlike standard procedures that are based only on the last measurement (or on moving average/variance, in which the number of historical values to include is determined arbitrarily). For this purpose, we analyzed data from a database of patients hospitalized with MI in Israel between 2002 and 2014.

The paper proceeds as follows: the next section describes the change-point model on which the study is based on; Section 3 presents a framework for analyzing potassium data of patients hospitalized with MI, for the purpose of detecting life-threatening signals; Section 4 describes the study population; Section 5 displays the results; Section 6 discusses substantial findings and draws conclusions.

## 2  The Change-point Model

The classical surveillance problem consists of being able to view sequentially a series of independent observations $X_1, X_2, X_3 ...$ such that $X_1, X_2, ..., X_{v-1}$ have distribution $F_0$ which changes at an unknown time $v$, so that $X_v, X_{v+1}, ...$ have distribution $F_1$. One applies a surveillance scheme that raises an alarm at time $N$, declaring that a change is in effect. A surveillance scheme is considered good if it detects a true change quickly, yet seldom raises a false alarm. The basic statistic in the suggested surveillance method is

$$\Lambda_k^n = \frac{f_{v=k}(X_1,...,X_n)}{f_{v=\infty}(X_1,...,X_n)} \qquad (1)$$

which is the likelihood ratio of the observations until time $n$, for $v = k$ (i.e. there was a change at time $k$) versus $v = \infty$ (i.e. there was no change until time $n$). Based on the Shiryaev-Roberts (SR) approach (Shiryaev, 1963; Roberts, 1966), the proposed method requires computing the sequence of statistics

$$R_n = \sum_{k=1}^n \Lambda_k^n \qquad (2)$$

and raising an alarm the first time that $R_n$ exceeds a threshold $A$; that is

$$N = N_A = \min \{n \mid R_n \geq A\}. \tag{3}$$

There are other surveillance procedures (in addition to SR) which can be applied in order to monitor the values of a process over time. Examples are Shewhart charts (Shewhart, 1931) and Cusum procedures (Page, 1954; Van Dobben de Bruyn, 1968). However, the SR approach has asymptotic optimality properties in terms of speed of detection (Pollak, 1985), and it can handle dependent data relatively easily.

Calculation of $\Lambda_k^n$ (as in Equation 1) assumes knowledge of the underlying distribution. Often, as in this case, both pre- and post-change parameters are unknown. Previously applied methods use a very large learning sample, which is equivalent to full information, but usually requires many observations to be practical, by which time a patient may have already died. Pollak and Siegmund (1991) suggested an approach to circumvent this type of problem. Their technique requires neither the complete information nor a large learning sample, thus we suggest to apply it here. Consider first a case where the observations are independent and normally distributed, but we do not know their mean nor their standard deviation. Namely, for each patient we observe a sequence of potassium values $X_1, X_2, \ldots$ where before a change $X_i \sim N(\mu_0, \tau^2)$, and we are concerned that this may change to a $N(\mu_0 + \delta\tau, \tau^2)$ or a $N(\mu_0 - \delta\tau, \tau^2)$ distribution (since both low and high potassium levels can be life-threatening). Neither $\mu_0$ nor $\tau$ is known, and the putative change is of $\delta$ standard deviations ($\delta$ can be regarded as a representative of the magnitude of a change that one would definitely want to detect, should it occur). In this paper we posit $\delta=1$; assuming any change in potassium mean is gradual, we do not expect to detect a very small change, yet we do expect to detect a change before the patient will die (if he/she will die).

Therefore, a change of one standard deviation seems reasonable as representative of what we wish to reveal. A more detailed discussion on $\delta$ appears in Pollak and Siegmund (1991).

Let $\bar{X}_n = \frac{1}{n}\sum_{i=1}^{n} X_i$. The sequence of standardized recursive residuals (Brown, Durbin and Evans, 1975) is defined as

$$Y_i = (X_i - \bar{X}_{i-1})\sqrt{\frac{i-1}{i}} \ ; \quad i = 2,3,\dots \qquad (4)$$

The distribution of the sequence of $Y_i$'s is independent of $\mu_0$. Now construct the sequence

$$Z_i = Y_i / Y_2 \ ; \quad i = 2,3,4,5,\dots \qquad (5)$$

The distribution of the sequence of $Z_i$'s is independent of both $\mu_0$ and $\tau$. Therefore, one can monitor the sequence of $Z_i$'s (instead of monitoring the process of $X_i$'s), where the pre-change and post-change densities are completely specified. It is therefore possible to compute the likelihood ratios $\Lambda_k^n$ for the $Z_i$ series; Pollak, Croarkin and Hagwood (1993) showed that for $n \geq k > 2$, the likelihood ratio of $Z_3, \dots, Z_n$ is

$$\Lambda_k^n = \frac{\int_{-\infty}^{\infty} |t + a_{k,n}|^{n-2} \exp(-0.5 t^2) dt}{\int_{-\infty}^{\infty} |t|^{n-2} \exp(-0.5 t^2) dt} \exp\left\{-0.5\delta^2 (k-1)^2 \left[\frac{1}{k-1} - \frac{1}{n}\right] + 0.5 a_{k,n}^2\right\} \qquad (6)$$

where

$$a_{k,n} = \frac{\delta(k-1) \sum_{i=k}^{n} \frac{Z_i}{\sqrt{i(i-1)}}}{\sqrt{\sum_{i=2}^{n} Z_i^2}} \ . \qquad (7)$$

(Although an expression for $k = 2$ can be obtained, we assume that if a change occurs, it does not occur within the first two observations, so that $k > 2$.) A computer program can calculate the sequence of $R_n$'s as defined by Equation (2) (see Appendix A for a program in MATLAB).

Obviously, there might be serial dependence between potassium measurements. In order to overcome this problem, we suggest to build an ARMA model (Box and Jenkins, 1970) for each patient separately, which best describes the dynamics of his/her potassium level, based on a learning period which is evidentially sufficient. This learning period should be as short as possible; the idea is to find the smallest number of measurements that yields a model which is similar to ARMA models built from a larger number of measurements. The estimated residuals of this ARMA model should be independent of each other with zero mean and constant variance (if the model is good). Moreover, if the mean of $X_i$ has actually changed, the mean of the residuals will also change, yet not in the same magnitude: if before a change $X_i \sim N(\mu_0, \tau^2)$, and after a change $X_i \sim N(\mu_0 + \delta\tau, \tau^2)$ and, for example, we have an AR($p$) process so that the model is $X_t = a_0 + a_1 X_{t-1} + \ldots + a_p X_{t-p} + \varepsilon_t$, where $\varepsilon_t \sim N(0, \sigma^2)$, then if a change occurred at $t = v$, then, for $t > v + p$, the mean of the residuals, $E(\varepsilon_t)$, equals $\delta\tau(1 - a_1 - \ldots - a_p)$; see proof in Appendix B. So, for $t > v + p$, $E(\varepsilon_t)$ is a constant. True, for $v < t < v + p$ it is not constant, but since $p$ is not large this will not make much of a difference. $E(\varepsilon_t) \neq 0$, since $(1 - a_1 - \ldots - a_p)$ must be greater than 0 for the model to remain stationary.

Thus, the residuals can be treated as $X_i$'s in Equation (4). Since we do not know what $\delta$ is, we should be looking for a two-sided change in the mean of the residuals. In addition, after fitting the AR($p$) model, one should hold in mind that if $(1 - \hat{a}_1 - \ldots - \hat{a}_p)$ is considerably smaller than 1, then the change in the residuals' mean is smaller than the change in the original $X_i$'s. Since the fitted ARMA model is, after all, only an approximation of reality, we suggest to assume no baseline is known (and thus not to posit that the residuals' mean is zero).

## 3 A Framework for monitoring of potassium levels in patients hospitalized for MI

Based on the change-point model presented in section 2, we formulated a framework for analyzing potassium data of patients hospitalized with MI, for the purpose of detecting life-threatening signals. The proposed methodology includes the following steps:

1. *Model identification*

The learning period, *m* (i.e., the number of measurements it takes to learn each patient's dynamics of potassium), needs to be specified first. It can be determined as the smallest number of measurements which yields a stable ARMA model, in the sense that it does not change much when more observations are considered. Once the learning period is specified, it is reasonable to assume that the first *m* potassium values constitute a stationary series. Therefore, the next step is to decide which autoregressive (AR) and/or moving average (MA) component should be used in the model. It should be noted here that in case of unequal time intervals between measurements, one needs to test first whether these intervals affect the potassium level. Formally, it means building an ARMA(*p*,*q*) model with an additional explanatory variable - the time between the present observation and the previous one $(t_i - t_{i-1})$. Namely:

$$X_{t_i} = a_0 + a_1 X_{t_{i-1}} + \cdots + a_p X_{t_{i-p}} + b_1 \varepsilon_{t_{i-1}} + \cdots + b_q \varepsilon_{t_{i-q}} + c_1(t_i - t_{i-1}) + \varepsilon_{t_i} \quad (8)$$

where $i = 1, \ldots, n$ and $X_{t_i}$ is the potassium level at time $t_i$. Estimating the parameters $a_0, \ldots, a_p, b_1, \ldots, b_p, c_1$ in Equation 8 will examine if there is no significant effect to the time between measurements (i.e. $c_1 \approx 0$ in Equation 8). If these time intervals do not have a significant effect, one may fit an ARMA model without regard to the differences between time intervals. (Otherwise, one needs to adjust the unevenly-spaced series by using one of the many approaches in this field.)

The next step is to arrive at coefficients that best fit the selected ARMA model. In this work we used the *auto.arima* function in package 'forecast' in R (Hyndman and Khandakar, 2008; Hyndman, 2015), as well as the *Expert Modeler* module in SPSS (release 20.0.0), which automatically find the best-fitting model in terms of AIC. Such modules exist in other statistical packages, yet using them is not mandatory; one may fit a number of ARMA models to the data, and use one or more measures to judge which is the best in terms of fit and parsimony.

2. *Model validation*

Validation is achieved by testing whether the residuals behave approximately like a white noise process. Commonly used tests are the Ljung-Box test (Ljung and Box, 1978) and the runs test (Wald and Wolfowitz, 1940). If it turns out that estimation is inadequate, it should be noted to the physician that a risk-control chart cannot be drawn yet.

3. *Construction of a control chart*

Once the appropriate model of the patient's potassium pattern is found, it should be applied to future measurements, in order to detect a change in the dynamics. That is, for each new potassium measurement, $X_n$ (starting from $n = m+1$), one needs to calculate the residual, $E_n = X_n - F_n$, where $F_n$ is the fitted value according to the model, and to compute $R_n$ (by using Equations 4-7 and 2 with $E_i$ replacing $X_i$) and plot the entire in-hospital control chart, namely all the points ($i$, $R_i$), for $i = 1,...,n$. The proposed algorithm will raise an alarm at observation $n=m$ if any of the $R_i$, $i = 1,…,m$, exceeds a specified threshold $A$ (eq.3), or else at observation $n>m$, where the value of $R_n$ first exceeds this threshold. Specifying $A$ can be done by plotting an ROC curve of a cohort of patients and choosing the threshold value that yields a satisfying combination of sensitivity and specificity.

## 4      Study Population

Patients included in this retrospective cohort study were hospitalized at Soroka university medical center from January 2002 through December 2014, as a result of MI (not necessarily the first). As it was found that the learning period needs to be of 40 measurements (to be explained in subsection 5.1), only patients with at least 40 potassium measurements were included (30 days in hospital on average). True, this leads to a bias towards a more at risk population, as such patients are usually severely ill. A fortiori, a method to identify those at high risk for arrhythmia and death among them is important, as they are considered to be clinically complicated and controlling their potassium level is clearly a challenge for cardiologists. In accordance with the aforementioned criteria, 195 patients were included in this research; 107 survived the MI and were discharged from hospital, 88 died at hospital. Potassium levels of these patients were measured every few hours; the average time between measurements was 18.3 hours, with SD of 24.7 hours (median = 11.3 hours, inter-quartile range = 9.4 hours). 29% of the survivors are females and 38% of non-survivors are females. A logistic regression was performed to ascertain the effects of age and gender on the likelihood that participants will die at hospital. The model was not statistically significant ($p = .826$). The mean entry age of survivors was 67.0 years (SD = 13.1) and the mean entry age of non-survivors was 71.4 years. Other demographics and co-morbidities are not in the scope of this study and thus have not been considered, yet are available upon request.

## 5   Results

The results are organized in accordance with the steps of the proposed framework (Section 3). As preliminary step, we applied the model from eq.8 on 20 random potassium series, to test whether the effect of the uneven time intervals is significant. For all the examined series, this effect was insignificant, thus we fitted ARMA models without regard to the differences between time intervals.

### 5.1   *The Learning Period (m)*

A random sample of 20 series was explored in order to specify the learning period. For 19 out of the 20 series, an ARMA model that was built from the first 40 observations was similar to ARMA models built from more observations (examining $n$ = 50, 60, 70, 80, 90, 100), both in ($p,q$) and in their estimates of parameters. In other words, $m = 40$ was found to be the smallest number of measurements that is reliable enough to learn the pattern of the potassium behavior.

### 5.2   *Fitting an ARMA model*

The next step was to test whether there is a serial dependence between measurements. According to the runs test, 58 out of 195 series were independent (testing the first 40 values). For the other 137 series, we found the ARMA model that best fits the data (in terms of AIC, individually for each patient) using the *Expert Modeler* module in SPSS as well as the *auto.arima* function in package 'forecast' in R. The two modules yielded similar models. An interesting finding was that for 132 series (out of 137), the best model was AR(1). For uniformity, we suggest to model all the series by an AR(1) model. To validate the models, we performed the Ljung-Box test and the runs test for independence of the residuals. Six series were not normally distributed, according to the Shapiro-Wilk test; for these series we used a nonparametric procedure for detecting a change in mean (Gordon and Pollak, 1994).

*5.3    The Threshold for Raising an Alarm (A)*

Once we found the appropriate model for each patient, we applied it to all his/her potassium measurements and calculated the residuals (in cases of serial dependence). The next step was to compute the sequence of statistics $R_1, \ldots, R_N$ (according to the MATLAB program given in appendix A). In order to choose a value of *A*, recall that in series that were modeled as AR(1), if $\hat{a}_1$ is close to 1, then the change in the residuals' mean is considerably smaller than the change in the original $X_i$'s. In such cases, the scale of $R_n$ of the post-change residuals will be much smaller than the scale of $R_n$, had the measurements been independent. Therefore, in this analysis (which yielded $\hat{a}_1$ ranging from 0.42 to 0.81), two different ROC curves should be drawn: one for patients whose measurements are independent and one for those whose residuals are AR(1). The ROC curves are drawn in Figure 1, based on the $maxR_n$ of each patient as test variable. Each curve is created by plotting the true positive rate (i.e., $sensitivity$) against the false positive rate (i.e., $1 - specificity$) at various threshold settings. Here, the true positive rate is the proportion of non-survivors for whom an alarm was raised (i.e., the conditional probability of raising an alarm given the patient will die), and the false positive rate is the proportion of survivors for whom an alarm was raised (erroneously). For both ROC curves that were drawn, the area under the curve was significantly different from 0.5 ($p = 0.01$ and $p < 0.001$, see

Appendix C), and therefore there is evidence that the $R_n$ criterion has the ability to distinguish between survivors and non-survivors.

There is no doubt that in this setting, the sensitivity of the test is of greater importance than its specificity. Therefore, only threshold values for which $sensitivity > specificity$ are considered (see the complete ROC tables in Appendix C). But how important is the sensitivity of the test, compared to its specificity? It is clearly hard to give an answer. For this reason, five weighted averages of sensitivity and specificity were considered:

$$g(A, \alpha) = \alpha \cdot Sensitivity(A) + (1 - \alpha) \cdot Specificity(A),$$

for $\alpha = .5, .6, .7, .8, .9$. Therefore, one should search for *A* that yields maximal values for each one of the five $g(A, \alpha)$'s. For patients whose measurements are independent, if $0.5 \leq \alpha \leq 0.7$ then the maximum of $g(A, \alpha)$ is obtained for *A*=674 (see Figure 2), which yields sensitivity of 81% and specificity of 56%; higher values of $\alpha$ produce unacceptable specificity (<10%). For patients whose measurements are not independent (i.e. residuals of *AR(1)*), if $\alpha = 0.5$ then $A = 101$, which yields sensitivity of 73% and specificity of 53%; if $\alpha = 0.6$ then $A = 73$, which yields sensitivity of 82% and specificity of 43% (see Figure 3); higher values of $\alpha$ produce unacceptable specificity (<10%); thus the choice between $A = 101$ and $A = 73$ depends on one's relative importance of sensitivity versus specificity (if they are considered equally important then $A = 101$).

*5.4    Average Run Length to False Alarm*

For $A = 101$ (an optimal threshold for patients whose measurements are not independent) and $\delta = 1$, the average run length (ARL) to false alarm, $E_\infty N$, is approximately 180 measurements (these values were predicted by asymptotic theory and estimated by monte carlo; cf. Pollak, 1987). When there is no change whatsoever, the run length (to false alarm) is approximately exponentially distributed (cf. Asmussen, 2003). Therefore, the probability that the run length would be less than 60, for example, is $1 - \exp(-60/180) = 28\%$, whereas the percentage of those who survived yet an alarm was raised within their first 60 measurements is 39%. For $A = 674$ and $\delta = 1$, the ARL is ≈1203 measurements, therefore the probability that the run length would be less than 60 is 5%, while the percentage of survivors for whom an alarm was raised within the first 60 measurements is 23%. The conclusion is that changes occur not only to non-survivors, but to survivors as well. However, changes in survivors' potassium level seem to be smaller, and thus it is possible to discriminate between survivors and non-survivors.

*5.5    Putting All Together*

Based on the results of this study, *if* such a proposal were to be applied (in hospitals), the following steps would have to be taken: (1) right after the 40$^{th}$ measurement, test whether these measurements are serially dependent; (2) if they are independent, compute $R_n$ (based on the original observations) for $n = 1, ... ,40$ and raise an alarm if $\max R_n > 674$; if $\max R_n < 674$, then for each new measurement $n>40$, compute $R_n$ and raise an alarm if $R_n > 674$. If the first 40 measurements are not independent, find the best AR(1) parameters for the data, compute the $R_n$ of the residuals for $n = 1, ... ,40$ and raise an alarm if $\max R_n > 101$; if $\max R_n < 101$, then for each new measurement $n>40$, compute its residual based on the original AR(1), calculate $R_n$ and raise an alarm if $R_n > 101$.

## 6    Discussion

We have presented an innovative application of a change detection method, for identifying hazardous changes in the serum potassium level of patients hospitalized due to a heart attack. This approach, in which changes in mean are suspected as harbingers of mortal danger, is novel to the problem of potassium-control in a few aspects; first, it considers all in-hospital potassium measurements of the patient; a change in mean may be masked by the variability of measurements, and only by considering a history of measurements (instead of only the last one) can it be detected early. Second, it is personalized to each patient, in the sense that it detects changes in his/her mean based on an individualized model, rather than crossing of absolute control limits. Third, no baseline knowledge of the patient is needed.

We presented evidence that the $R_n$ criterion has a significant ability to discriminate between survivors and non-survivors in our data. This raised the problem of when to raise an alarm (i.e., for which threshold value of $R_n$), where a trigger-happy protocol may result in too many false alarms but a timid approach may result in a late detection. We therefore probed the consequences of various choices of parameters, so the method detected approximately 80% of the patients that eventually died in our cohort, while wrongly warning for only ~40% of the survivors.

The proposed method is not intended for replacing approaches that monitor the crossing of an upper or lower absolute boundary; the relative importance of absolute level versus change in mean needs to be considered before proposing a change in clinical practice. Furthermore, if we had known which treatment had been given to each patient, perhaps we could have improved the detection procedure. However, our analysis aims to show that the 'change in mean' approach has considerable power to predict whether the patient is at risk of in-hospital death. To summarize, we present our analysis as evidence that the proposed approach has a significant added value, in addition to the standard protocol.

In addition to all the above, there exist many other medical circumstances where a certain variable of a patient is monitored on a daily basis, with the goal of discerning changes that indicate increased risk (e.g., glucose level of diabetics is measured a few times a day). Application of a method as suggested here on such stochastic processes may possibly prevent severe scenarios that are associated with changes in these variables.

**Acknowledgements**

This research was supported by the Marcy Bogen Chair of Statistics and by Israel National Science Foundation grant no. 1450/13.

# Appendix A. MATLAB program for computing $R_n$ (Eq.2) for a two-sided SR scheme detecting a change in a normal mean with unknown initial mean and variance

```
% Input parameters: data (row vector of size en)
%                   d (delta, the representation of the change in the mean)
%Output:            R (row vector of size en, giving the values of the Shiryaev-
%                   Roberts statistic for n=1:en)

en=length(data);
X=cumsum(data);
R=zeros(1,en);
Y=R;
W=R;
R(1)=1;
R(2)=2;
for n=3:en
    t=zeros(1,n);
    t(1)=1;
    for i=2:n
        Y(i)=(data(i)-(X(i-1)/(i-1)))*sqrt((i-1)/i);
        W(i)=Y(i)/sqrt(i*(i-1));
    end
    s=sqrt(Y*Y');
    W=W(:,n:-1:1);
    W=cumsum(W);
    W=W(:,n:-1:1);
    x=ones(1,n);
    x=cumsum(x)-1;
    a=(d/s)*x.*W;
    u=zeros(1,n);
    v=u;
    w=u;
    y=u;
    for k=2:n
        p=(exp((-a(k)^2)/2))/2;
        q=(erf(a(k)/sqrt(2)))/2+0.5;
        u(1)=p-a(k)*(1-q)*sqrt(pi/2);
        v(1)=p+a(k)*q*sqrt(pi/2);
        w(1)=u(1)*sqrt(2/pi);
        y(1)=v(1)*sqrt(2/pi);
        u(2)=1-q-a(k)*w(1);
        v(2)=q+a(k)*y(1);
        w(2)=(1-q)*sqrt(pi/2)-a(k)*u(1);
        y(2)=q*sqrt(pi/2)+a(k)*v(1);
        for j=3:n
            u(j)=u(j-2)-a(k)*w(j-1)/(j-1);
            w(j)=((j-1)/(j-2))*w(j-2)-a(k)*u(j-1);
            v(j)=v(j-2)+a(k)*y(j-1)/(j-1);
            y(j)=((j-1)/(j-2))*y(j-2)+a(k)*v(j-1);
        end
        t(k)=(u(n-2)+v(n-2))*exp(-0.5*((d*(k-1))^2)*((1/(k-1))-(1/n)-
    (W(k)/s)^2));
        t(2)=t(2)*exp(-0.25*(d^2));
    end
    R(n)=sum(t);
end
```

# Appendix B.

***Proposition.*** Suppose we have an ARMA($p,q$) process so that the model is

$$X_t = a_0 + a_1 X_{t-1} + \ldots + a_p X_{t-p} + b_1 \varepsilon_{t-1} + \cdots + b_q \varepsilon_{t-q} + \varepsilon_t, \quad (B1)$$

before a change $X_i \sim N(\mu_0, \tau^2)$, after a change $X_i \sim N(\mu_0 + \delta\tau, \tau^2)$ and the change occurred at $t = \nu$. Then, for $t > \nu + max(p,q)$, the mean of the residuals is: $E(\varepsilon_t) = \frac{\delta\tau(1-a_1-\ldots-a_p)}{1+b_1+\cdots+b_q}$.

***Proof.*** Pre-change, the model is stationary, with $E(X_t)$ identical for all values of $t < \nu$. In addition, $\varepsilon_t$ is defined as $N(0, \sigma^2)$. Therefore, it is easy to show (from eq.B1) that

$$\mu_0 = a_0/(1 - a_1 - \ldots - a_p). \quad (B2)$$

If a change occurred at $t = \nu$, so that $E(X_i) = \mu_0 + \delta\tau$ for $i \geq \nu$, and $E(\varepsilon_i)$ is identical for all $i \geq \nu$, then, for $t > \nu + max(p,q)$, we get (from eq.B1 and eq.B2):

$$E[b_1\varepsilon_{t-1} + \cdots + b_q\varepsilon_{t-q} + \varepsilon_t] = E[X_t - (a_0 + a_1 X_{t-1} + \ldots + a_p X_{t-p})]$$

$$\Rightarrow E(\varepsilon_t) \cdot (1 + b_1 + \cdots + b_q) = \mu_0 + \delta\tau - [a_0 + a_1(\mu_0 + \delta\tau) + \ldots + a_p(\mu_0 + \delta\tau)]$$

$$\Rightarrow E(\varepsilon_t) \cdot (1 + b_1 + \cdots + b_q) = \mu_0(1 - a_1 - \ldots - a_p) - a_0 + \delta\tau(1 - a_1 - \ldots - a_p)$$

$$\Rightarrow E(\varepsilon_t) \cdot (1 + b_1 + \cdots + b_q) = a_0 - a_0 + \delta\tau(1 - a_1 - \ldots - a_p)$$

$$\Rightarrow E(\varepsilon_t) = \frac{\delta\tau(1-a_1-\ldots-a_p)}{1+b_1+\cdots+b_q}.$$

# Appendix C. Coordinates of the ROC Curves and Statistics on the area under the curves

## (1)   ROC of the originally-independent series

|  | N |
|---|---|
| Died | 26 |
| Survived | 32 |

**Area Under the Curve**
Test Result Variable: Max_Rn

| Area | Std. Error[a] | Asymptotic Sig.[b] | Asymptotic 95% Confidence Interval | |
|---|---|---|---|---|
| | | | Lower Bound | Upper Bound |
| .698 | .071 | .010 | .559 | .838 |

a. Under the nonparametric assumption
b. Null hypothesis: true area = 0.5

| Positive if Greater Than or Equal To[a] | Sensitivity | 1 - Specificity |
|---|---|---|
| 27 | 1.00 | 1.00 |
| 32 | 1.00 | 0.97 |
| 43 | 1.00 | 0.94 |
| 51 | 1.00 | 0.91 |
| 65 | 0.96 | 0.91 |
| 84 | 0.92 | 0.91 |
| 93 | 0.88 | 0.91 |
| 114 | 0.88 | 0.88 |
| 138 | 0.88 | 0.84 |
| 154 | 0.88 | 0.81 |
| 179 | 0.88 | 0.78 |
| 203 | 0.88 | 0.75 |
| 248 | 0.88 | 0.72 |
| 292 | 0.85 | 0.72 |
| 304 | 0.85 | 0.69 |
| 330 | 0.81 | 0.69 |
| 366 | 0.81 | 0.66 |
| 380 | 0.81 | 0.63 |
| 381 | 0.81 | 0.59 |
| 475 | 0.81 | 0.56 |
| 577 | 0.81 | 0.53 |
| 591 | 0.81 | 0.50 |
| 630 | 0.81 | 0.47 |
| 674 | 0.81 | 0.44 |
| 700 | 0.77 | 0.44 |
| 768 | 0.73 | 0.44 |
| 1106 | 0.73 | 0.41 |
| 1657 | 0.73 | 0.38 |
| 2021 | 0.69 | 0.38 |
| 2538 | 0.69 | 0.34 |
| 3172 | 0.65 | 0.34 |
| 3387 | 0.62 | 0.34 |
| 3605 | 0.58 | 0.34 |
| 4018 | 0.54 | 0.34 |
| 4251 | 0.50 | 0.34 |
| 4607 | 0.50 | 0.31 |
| 4958 | 0.50 | 0.28 |
| 5235 | 0.50 | 0.25 |
| 5814 | 0.46 | 0.25 |
| 7714 | 0.42 | 0.25 |
| 10164 | 0.42 | 0.22 |
| 12440 | 0.42 | 0.19 |
| 14332 | 0.42 | 0.16 |
| 16954 | 0.42 | 0.13 |
| 19690 | 0.38 | 0.13 |
| 25387 | 0.38 | 0.09 |
| 30879 | 0.38 | 0.06 |
| 33277 | 0.35 | 0.06 |
| 48598 | 0.31 | 0.06 |
| 130383 | 0.31 | 0.03 |

( a. The smallest cutoff value is the minimum observed test value minus 1, and the largest cutoff value is the maximum observed test value plus 1. All the other cutoff values are the averages of two consecutive ordered observed test values.)

## (2) ROC of AR(1) residuals (of originally- dependent series)

|  | N |
|---|---|
| Died | 62 |
| Survived | 75 |

**Area Under the Curve**

Test Result Variable: max_Rn

| Area | Std. Error[a] | Asymptotic Sig.[b] | Asymptotic 95% Confidence Interval | |
|---|---|---|---|---|
| | | | Lower Bound | Upper Bound |
| .692 | .046 | .000 | .602 | .783 |

a. Under the nonparametric assumption
b. Null hypothesis: true area = 0.5

| Positive if Greater Than or Equal To[a] | Sensitivity | 1 - Specificity |
|---|---|---|
| 8 | 1 | 1 |
| 10 | 1 | 0.987 |
| 11.5 | 1 | 0.973 |
| 15.5 | 0.984 | 0.973 |
| 19.5 | 0.984 | 0.96 |
| 20.5 | 0.968 | 0.947 |
| 21.5 | 0.952 | 0.947 |
| 22.5 | 0.952 | 0.933 |
| 24.5 | 0.952 | 0.92 |
| 26.5 | 0.952 | 0.907 |
| 28 | 0.935 | 0.907 |
| 29.5 | 0.935 | 0.893 |
| 32 | 0.935 | 0.88 |
| 35.5 | 0.919 | 0.867 |
| 37.5 | 0.919 | 0.853 |
| 38.5 | 0.919 | 0.84 |
| 39.5 | 0.919 | 0.827 |
| 40.5 | 0.919 | 0.813 |
| 42.5 | 0.903 | 0.813 |
| 44.5 | 0.903 | 0.8 |
| 45.5 | 0.903 | 0.787 |
| 47 | 0.887 | 0.787 |
| 49.5 | 0.871 | 0.773 |
| 53 | 0.871 | 0.76 |
| 55.5 | 0.855 | 0.733 |
| 57.5 | 0.855 | 0.72 |
| 60.5 | 0.855 | 0.693 |
| 62.5 | 0.855 | 0.64 |
| 63.5 | 0.839 | 0.64 |
| 65.5 | 0.823 | 0.627 |
| 69.5 | 0.823 | 0.587 |
| 73 | 0.823 | 0.573 |
| 75.5 | 0.806 | 0.573 |
| 78 | 0.79 | 0.547 |
| 80 | 0.758 | 0.547 |
| 81.5 | 0.758 | 0.533 |
| 84 | 0.758 | 0.52 |
| 86.5 | 0.758 | 0.507 |
| 88.5 | 0.742 | 0.507 |
| 92 | 0.742 | 0.493 |
| 95.5 | 0.726 | 0.493 |
| 97.5 | 0.726 | 0.48 |
| 101 | 0.726 | 0.467 |
| 104.5 | 0.694 | 0.467 |
| 105.5 | 0.694 | 0.453 |
| 106.5 | 0.677 | 0.453 |
| 107.5 | 0.677 | 0.44 |
| 108.5 | 0.661 | 0.44 |
| 114 | 0.661 | 0.427 |
| 123 | 0.645 | 0.427 |
| 127.5 | 0.629 | 0.413 |
| 128.5 | 0.629 | 0.4 |
| 129.5 | 0.629 | 0.387 |
| 132.5 | 0.629 | 0.373 |
| 135.5 | 0.613 | 0.373 |
| 137.5 | 0.613 | 0.36 |
| 143.5 | 0.597 | 0.347 |
| 156.5 | 0.597 | 0.333 |
| 169 | 0.597 | 0.32 |
| 174.5 | 0.581 | 0.32 |
| 177.5 | 0.581 | 0.28 |
| 180.5 | 0.581 | 0.267 |
| 189.5 | 0.581 | 0.253 |
| 198.5 | 0.581 | 0.227 |
| 201 | 0.581 | 0.2 |
| 205 | 0.581 | 0.187 |
| 214 | 0.581 | 0.173 |
| 228.5 | 0.565 | 0.173 |
| 242 | 0.548 | 0.173 |
| 249 | 0.532 | 0.173 |
| 251.5 | 0.532 | 0.16 |
| 256 | 0.516 | 0.16 |
| 262 | 0.5 | 0.16 |
| 271 | 0.484 | 0.16 |
| 290 | 0.484 | 0.147 |
| 307 | 0.468 | 0.147 |
| 325.5 | 0.468 | 0.133 |
| 342.5 | 0.452 | 0.133 |
| 354 | 0.435 | 0.133 |
| 363 | 0.419 | 0.133 |
| 373.5 | 0.403 | 0.133 |
| 388.5 | 0.403 | 0.12 |
| 404 | 0.403 | 0.107 |
| 423.5 | 0.403 | 0.093 |
| 484 | 0.387 | 0.093 |
| 559 | 0.371 | 0.093 |
| 613 | 0.355 | 0.093 |
| 672 | 0.339 | 0.093 |
| 762 | 0.323 | 0.093 |
| 900 | 0.306 | 0.093 |
| 1021.5 | 0.29 | 0.093 |
| 1094 | 0.29 | 0.08 |
| 1126.5 | 0.274 | 0.08 |
| 1163.5 | 0.274 | 0.067 |
| 1201.5 | 0.258 | 0.067 |
| 1270 | 0.242 | 0.067 |
| 1431.5 | 0.242 | 0.053 |
| 2019 | 0.226 | 0.053 |
| 2682 | 0.21 | 0.053 |
| 4860.5 | 0.21 | 0.04 |
| 8108.5 | 0.194 | 0.04 |
| 9908 | 0.177 | 0.04 |
| 11947.5 | 0.161 | 0.04 |
| 37447.5 | 0.145 | 0.04 |
| 66041 | 0.129 | 0.04 |
| 72033 | 0.113 | 0.04 |
| 74887.5 | 0.097 | 0.04 |
| 90661 | 0.081 | 0.04 |
| 333033.5 | 0.081 | 0.027 |
| 1531681 | 0.065 | 0.027 |
| 3532800 | 0.048 | 0.027 |
| 7671029 | 0.032 | 0.027 |
| 20656921 | 0.032 | 0.013 |

( a. The smallest cutoff value is the minimum observed test value minus 1, and the largest cutoff value is the maximum observed test value plus 1. All the other cutoff values are the averages of two consecutive ordered observed test values.)

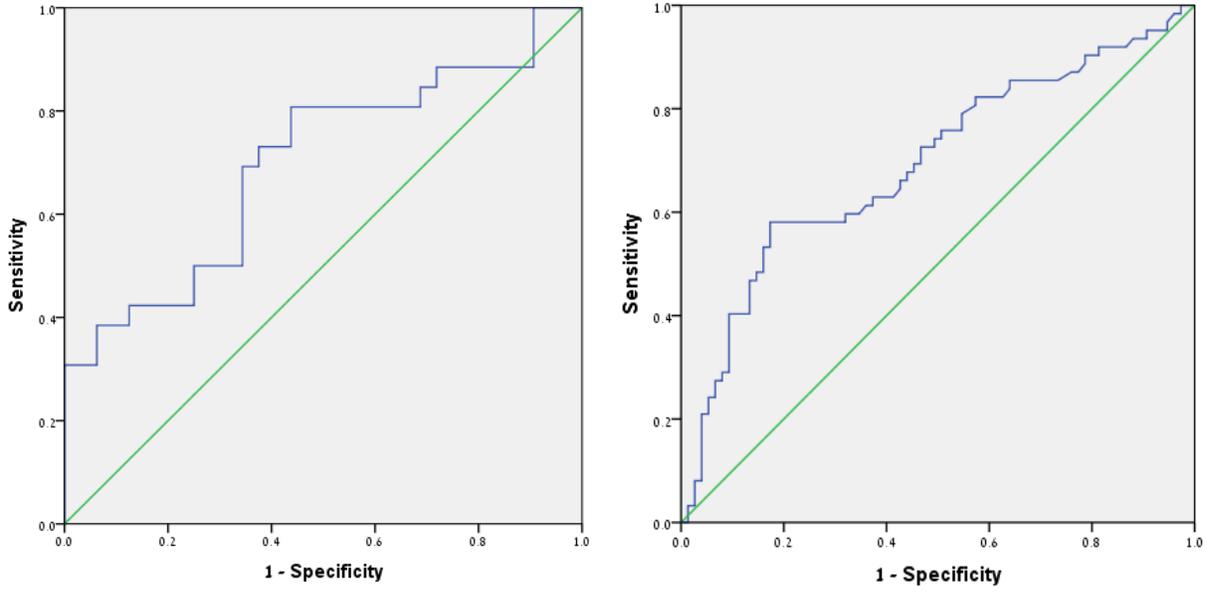

**Figure 1.** ROC curves: the left is of originally-independent series, the right is of AR(1) residuals (of originally-dependent series)

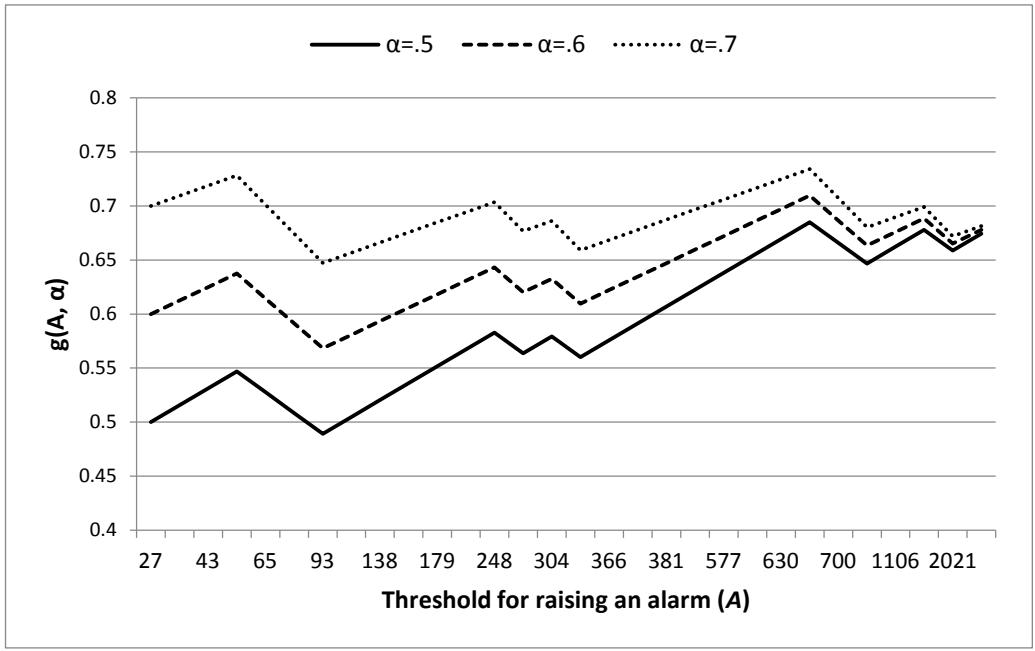

**Figure 2.** Weighted average, $g(A, \alpha)$, of sensitivity and specificity in the ROC of (originally) independent series, for three values of α (the weight of sensitivity).

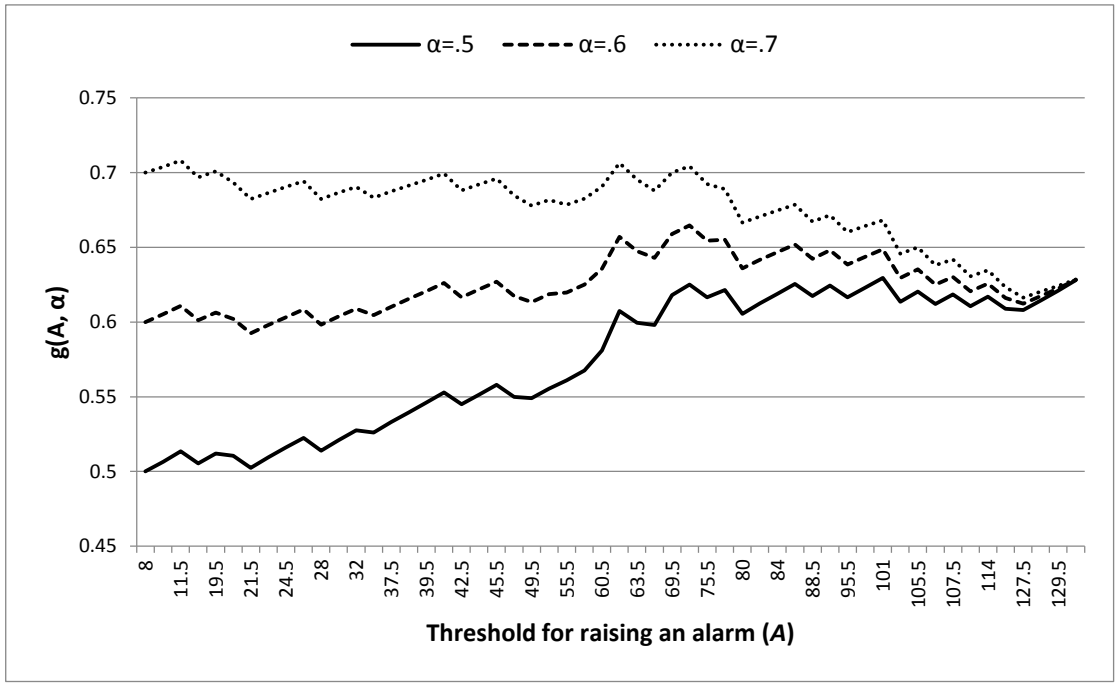

**Figure 3.** Weighted average, $g(A, \alpha)$, of sensitivity and specificity in the ROC of (originally) dependent series, for three values of α (the weight of sensitivity).